\documentclass[runningheads]{llncs}
\usepackage{citesort}
\usepackage{graphicx}
\usepackage{booktabs}
\usepackage{fontawesome}
\usepackage{amsmath}
\usepackage{url}
\usepackage{tikz}
\usepackage[strings]{underscore}
\begin{document}
\title{Defining Utility Functions for\\ Multi-Stakeholder Self-Adaptive Systems}

\author{Rebekka Wohlrab\textsuperscript{\faEnvelopeO}\orcidID{0000-0002-5449-7900} \and
	David Garlan\orcidID{0000-0002-6735-8301}}
\institute{School of Computer Science, Carnegie Mellon University, USA\\
\email{wohlrab@cmu.edu, garlan@cs.cmu.edu}}
\authorrunning{Rebekka Wohlrab and David Garlan}

\maketitle
\begin{abstract}
	\vspace{-1em}
		\begin{tikzpicture}[overlay]
		\node[draw, align=left, fill=white, thick] (b) at (3.5,6){ This is the Author’s Accepted Manuscript of an article published in REFSQ 2021:\\ Requirements Engineering: Foundation for Software Quality. The final version is available\\ via Springer at: \url{https://doi.org/10.1007/978-3-030-73128-1_8}};
	\end{tikzpicture}
	\textbf{[Context and motivation:]} For realistic self-adaptive systems, multiple quality attributes need to be considered and traded off against each other.
	These quality attributes are commonly encoded in a \textit{utility function}, for instance, a weighted sum of relevant objectives.
	\textbf{[Question/problem:]} The research agenda for requirements engineering for self-adaptive systems has raised the need for decision-making techniques that consider the trade-offs and priorities of multiple objectives.
	Human stakeholders need to be engaged in the decision-making process so that the relative importance of each objective can be correctly elicited.
	\textbf{[Principal ideas/results:]} This research preview paper presents a method that supports multiple stakeholders in prioritizing relevant quality attributes, negotiating priorities to reach an agreement, and giving input to define utility functions for self-adaptive systems.
	\textbf{[Contribution:]} The proposed method constitutes a lightweight solution for utility function definition. It can be applied by practitioners and researchers who aim to develop self-adaptive systems that meet stakeholders' requirements. We present details of our plan to study the application of our method using a case study.
	\keywords{self-adaptive systems \and quality attributes \and utility functions \and Analytic Hierarchy Process.}
\end{abstract}
\section{Introduction} \label{sec:introduction}
For self-adaptive systems, multiple quality attributes (such as performance, availability, and security) need to be considered and traded off against each other.
These quality attributes are often encoded in a \textit{utility function}, i.e., a single aggregate function whose expected value should be maximized by the system~\cite{Sawyer2010,Heaven2009,Inverardi2013,Faniyi2014}.
In self-adaptive systems, utility functions are typically used by automated planning mechanisms to identify the relative costs and benefits of alternative strategies.
In related work, utility functions are often defined as the weighted sum of relevant objectives~\cite{Ghezzi2013,Cheng2006,Sousa2008}.
For most approaches using utility functions, it is simply stated that they should be manually defined, but little guidance for this task is provided~\cite{Inverardi2013,Sousa2008}.
It is challenging to correctly identify the weights of each objective and consider trade-offs between multiple quality attributes, as reported in the research agenda for requirements engineering for self-adaptive systems~\cite{Sawyer2010}.
Self-adaptive systems often have multiple stakeholders (e.g., end users or business owners) whose preferences need to be consolidated to identify the overall relative importance of each objective~\cite{Salehie2012}.
Decision-making techniques are needed to help stakeholders prioritize and negotiate quality attributes and determine appropriate utility function weights~\cite{Sawyer2010}.

In this paper, we present a lightweight tool-supported method for utility function definition for multi-stakeholder self-adaptive systems.
The proposed method is based on the Analytic Hierarchy Process (AHP)~\cite{Saaty1987} and the Delphi technique~\cite{Hsu2007}.
It supports stakeholders in prioritizing quality attributes, negotiating priorities to reach an agreement, recording rationales and comments, and giving input to define utility functions.
{We use the weighted sum approach for utility functions, as it is lightweight and commonly used in related work}~\cite{Ghezzi2013,Cheng2006,Sousa2008}.
{It assumes that the weighted quantity of one quality attribute can be traded off (or ``substituted''~\cite{Abdennadher2018}) with another one.}
{Guidelines to select utility functions, considering risk and dependencies between decision parameters, have been previously created~\cite{Abdennadher2018} and can be used to support other kinds of utility functions in the future.}
The proposed ideas of this research preview paper will be further refined and we plan to study the method's application in a case study.
We expect that our method will be of use to practitioners and researchers that aim to conceive self-adaptive systems fulfilling stakeholders' preferences.
\section{Proposed Approach} \label{sec:proposed_approach}
\begin{figure}[b]
	\vspace{-0.5em}
	\centering
	\includegraphics[width=\linewidth]{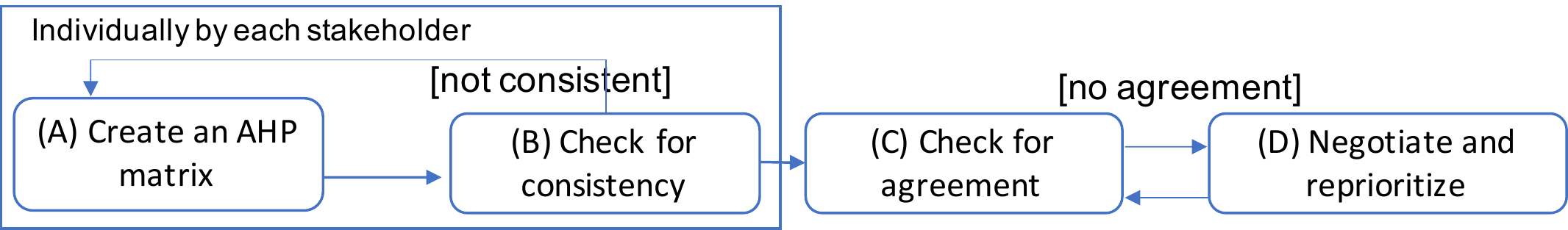}
	\caption{Overview of our method for utility function definition}
	\label{fig:ahputility}
	\vspace{-1.7em}
\end{figure}

Figure~\ref{fig:ahputility} shows the steps of our method for utility function definition.
The method can either be used for the initial definition or the refinement of the utility function, in case stakeholders' preferences evolve over time.
The two leftmost steps are performed individually by each stakeholder.
The two guard conditions refer to whether an AHP matrix is inconsistent and whether no agreement has been reached.
Each step is labeled with the paragraph in which it is described.

\vspace{0.2em}
\noindent{\textbf{(A) Create an AHP Matrix:}} \label{sec:createAHPMatrix}
For the prioritization of quality attributes, we use the AHP, which is especially useful when subjective, abstract, or non-quantifiable criteria are relevant for a decision~\cite{Saaty1987}.
A central part of the AHP is to elicit stakeholders' priorities of different objectives in pairwise comparison matrices, which are positive and reciprocal (i.e., $a_{ij} = 1/a_{ji}$).
For utility functions, we are interested in the degree of preference of one quality attribute over another, with the goal of increasing the overall utility of a system.
Verbal expressions are used for these pairwise comparisons (e.g., ``\textit{I strongly prefer X over Y}'').
Table~\ref{tab:numericalscale} shows how the verbal expressions correspond to numerical values.

For a robot planning problem, Table~\ref{tab:ahpexample} shows an example of an AHP matrix with the attributes safety (expected number of collisions), speed (duration of a mission), and energy consumption (consumed watt-hours).
In the example, safety is \textit{very strongly} preferred over speed (7) and \textit{extremely preferred} over energy consumption (9).
Speed and energy consumption are \textit{equally preferred}.

The relative priorities of the quality attributes can then be calculated using the principal eigenvector of the eigenvalue problem $Aw = \lambda_{max}w$~\cite{Saaty1987}.
$A$ is the matrix of judgments and $\lambda_{max}$ is the principal eigenvalue.
For the matrix in Table~\ref{tab:ahpexample}, the principal eigenvalue is $\lambda_{max}\approx 3.01$.
A corresponding normalized eigenvector to $\lambda_{max}$ is $(0.8, 0.1, 0.1)^T$, which corresponds to the relative priorities of the quality attributes.
The utility function for a mission might be defined as $U(m) = 0.8 \cdot \textsf{safety}(m) + 0.1 \cdot \textsf{duration}(m) + 0.1 \cdot \textsf{energy}(m)$.
$\textsf{safety}(m)$ indicates the expected number of collisions in a mission, $\textsf{duration}(m)$ the number of timesteps, and $\textsf{energy}(m)$ the consumed watt-hours.
The preference of a quality attribute can often be described with a sigmoid function defining an interval for the quantity that is considered as good enough and an interval for the quantity that is insufficient~\cite{Poladian2004}.
Appropriate methods will need to be created in the future to elicit these thresholds and define quality attributes' preference functions.
\begin{table} [b]
	\begin{footnotesize}
	\vspace{-1em}
\parbox{.45\linewidth}{
	\caption{AHP judgment/preference options with numerical values~\cite{Saaty1987}.}
	\label{tab:numericalscale}
	\vspace{-0.85em}
	\begin{tabular}{@{}p{.35\textwidth}p{.1\textwidth}@{}}
		\toprule
		Extremely preferred & 9 \\ 
		Very strongly preferred & 7 \\ 
		Strongly preferred & 5 \\ 
		Moderately preferred & 3 \\ 
		Equally preferred & 1 \\ 
		Intermediate values & 2, 4, 6, 8 \\
		\bottomrule
\end{tabular}} \hspace{0.4em}
\parbox{.45\linewidth}{
	\caption{Example of an AHP matrix.}
	\label{tab:ahpexample}
	\begin{tabular}{@{}p{.17\textwidth}@{}p{.08\textwidth}p{.08\textwidth}p{.17\textwidth}@{}}
		\toprule
		& Safety & Speed & Energy\newline Consumption \\ \midrule
		Safety & 1 & 7 & 9 \\ \midrule
		Speed & $\frac{1}{7}$ & 1 & 1 \\ \midrule
		Energy Cons. & $\frac{1}{9}$ & 1 & 1 \\ \bottomrule
\end{tabular}}
\vspace{-0.5em}
	\end{footnotesize}
\end{table}

\vspace{0.2em}
\noindent{\textbf{(B) Check for Consistency:}}
AHP matrices can be checked for consistency.
A matrix is consistent if $a_{jk} = a_{ik}/a_{ij}$ for $i,j,k = 1, \dots, n$~\cite{Saaty1987}.
Saaty proved that a necessary and sufficient condition for consistency is that the principal eigenvalue of A be equal to $n$, the order of A~\cite{Saaty1987}.
He defined the consistency index CI as $(\lambda_{max}-n)/(n-1)$.
For our example in Section~\ref{sec:createAHPMatrix}, CI is $0.004$.
To compare consistency values, Saaty also calculated the \textit{random consistency index} RI by calculating CI for a large number of reciprocal matrices with random entries~\cite{Saaty1987}.
For a $3\times3$ matrix, the average random consistency index was $0.58$.
According to Saaty, the consistency ratio $CR = CI/RI$ shall be less or equal to 0.10 for the matrix to be considered consistent~\cite{Saaty1987}.
In our example, the consistency ratio is 0.01.
If consistency is not fulfilled, stakeholders are required to refine their AHP matrices.
The matrix can be automatically analyzed to point out the triples of quality attributes QA$_{i}$, QA$_{j}$, and QA$_{k}$ where $a_{jk} \ll a_{ik}/a_{ij}$ or $a_{jk} \gg a_{ik}/a_{ij}$.

\vspace{0.2em}
\noindent{\textbf{(C) Check for Agreement:}}
We consider the rankings of $n$ quality attributes by $k$ stakeholders (where each quality attribute's rank is a number between 1 and $n$).
For QA$_i$, the sum of ranks by all stakeholders is $R_i$, and the mean value of these ranks is $\bar{R} = \frac{1}{n}\sum_{i=1}^{n} R_i$.
If the stakeholders' rankings do not agree, we can assume that the sums of ranks of several quality attributes are approximately equal~\cite{Kendall1939}.
It is therefore natural to consider the sum of squared deviations from the mean values of ranks $S = \sum_{i=1}^{n} (R_i - \bar{R})^2$~\cite{Kendall1939}.
The maximum possible value of S is $k^2(n^3 - n)/12$~\cite{Kendall1939}.
Kendall's concordance coefficient, describing the agreement of rankings in a [0,1] interval, is therefore: $W = \frac{12S}{k^2 \cdot (n^3 - n)}$~\cite{Kendall1939}.

\vspace{0.2em}
\noindent{\textbf{(D) Negotiate and Reprioritize:}}
In case agreement is not reached, a tool-supported negotiation and reprioritization phase starts.
To aggregate AHP matrices, the ``most recommendable aggregation technique'' is to calculate the weighted arithmetic mean of individual priorities (AIP)~\cite{Ossadnik2016}.
A priority indicates the importance of a quality attribute with a value between 0 and 1.
Stakeholders' priorities can be weighted differently, as their influence and stake may differ.

While the AIP can be used to quickly arrive at a solution, it is beneficial to discuss and record underlying rationales.
We adapt the \textit{Delphi technique}~\cite{Hsu2007} for remote consensus building.
Interactive tooling is used to support the technique and collect data.
The stakeholders anonymously provide input in several iterations and receive controlled feedback.
Users can declare that they do not know or do not care about a quality attribute.
It is also possible to delegate votes to another participant (proxy voting).
In the first round, open-ended questions are used concerning participants' rationales (e.g., ``in what situation(s) do you think safety is especially important?'').
The answer is fed back to other participants to inform their rankings.
The main trade-offs and conflicts between quality attributes are elicited and discussed.
While we assume an existing set of quality attributes, participants can also suggest new quality attributes and objectives.

In the second round, the comments and rationales are presented to the participants and the AHP matrices can be revised.
The rankings that are in conflict are indicated to increase transparency.
Further comments and rationales are added and a consensus starts to form.
In the third round, participants are asked to revise their judgments or declare why they decide to remain outside the consensus~\cite{Hsu2007}.
The final utility function is a weighted sum of the objectives, where the final weights are the participants' aggregated weighted priorities (using AIP).
\vspace{-1.1em}
\section{Empirical Study} \label{sec:evaluation}
\vspace{-0.6em}
We plan to perform a multiple case study~\cite{Runeson2009} focusing on the phenomenon of applying our approach in practice.
The approach depends on contextual factors and is therefore difficult to study in controlled settings (e.g., experiments).
As mentioned before, utility functions are a central mechanism in several approaches for self-adaptive systems.
We plan to apply the method to existing systems and projects.
As the first case, we focus on robot mission planning using a probabilistic model checker, where the correct definition of the weights of multiple objectives is essential.
The participants operate from multiple locations and are aware of the system's context and a preliminary set of quality attributes.
The stakeholders have conflicting objectives (e.g., end user, business/cost, performance, and safety concerns) and are asked to apply our method for utility definition.
We aim to use collected tool data, observations, and complementary interviews to study the required time to build a consensus, the understandability of the approach, as well as negotiation strategies.
The empirical study is intended to give insights into how participants typically act to reach a consensus, how beneficial our proposed method is perceived for utility function definition, and how satisfied stakeholders are with the resulting utility function.
\vspace{-1em}
\section{Related Work} \label{sec:related_work}
\vspace{-0.5em}
Identifying and prioritizing objectives for self-adaptive systems is a nontrivial task.
The Goal-Action-Attribute Model requires the goals, priorities, and preferences of multiple stakeholders to be elicited~\cite{Salehie2012}.
The AHP is suggested to be used for this task, but no concrete guidance is given.
We focus on prioritization and negotiation and develop a comprehensive AHP-based method.
Rather than focusing on creating complete goal models, we aim to create a lightweight method for utility function definition.
For security requirements, the Swing-Weight Method has been used for prioritization and utility function definition~\cite{Butler2002}.
The AHP allows a more precise elicitation of the relative priorities of objectives.

Utility functions are a common mechanism in self-adaptive systems~\cite{Heaven2009,Ghezzi2013,Faniyi2014,Cheng2006,Sousa2008}.
A few approaches for utility function definition are related to our work on the prioritization and negotiation of utility function weights.
Song et al.~\cite{Song2013} propose to collect user feedback after every round of adaptation to adjust the weights of constraints.
Another approach relies on user feedback to switch between ``variants'' with associated utility function weights, depending on the current usage context~\cite{Kakousis2008}.
Our work focuses on eliciting priorities to define utility function weights based on a consensus between multiple stakeholders.
As part of future work, we aim to also consider different usage contexts/scenarios in our method.
\vspace{-1.9em}
\section{Discussion, Conclusion, and Future Work} \label{sec:discussion}
\vspace{-0.7em}
In this paper, we presented a method to define utility functions for self-adaptive systems by eliciting and negotiating the priorities of quality attributes.
The method is based on the AHP for the pairwise comparison of quality attributes and a consensus-building approach using the Delphi technique.
We plan to study the method's application on existing systems in a multiple case study.
The method is at an early stage of investigation and needs to be refined further.
For instance, the current method assumes that stakeholders are aware of relevant and measurable quality attributes that can be expressed in functions.
For individual quality attributes, techniques are needed to define the intervals of values that are considered ``good enough'' or ``insufficient.''
Our method will also be extended to explicitly consider hard constraints.
Criteria mandated by law (e.g., security or safety constraints) cannot be traded against other preferences.
Moreover, the utility of a system strongly depends on its context (e.g., current tasks, security attacks, or faults), which should also be considered, so that human input for utility function definition can be collected when needed and the utility function can be evolved over time.
Another relevant concern is to ensure that stakeholders do not over-rate quality attributes to counter for others' conflicting preferences, as it is not always possible to assume non-competitive stakeholders.

We envision our method to be integrated into existing approaches, so that multiple stakeholders' preferences and requirements can be more easily elicited, negotiated, and fulfilled.
The presented ideas might also be beneficial for artifacts at other levels of abstraction, e.g., the prioritization of goals or requirements.
Moreover, we plan to work on explaining utility functions by describing different priorities' impact on the concrete actions of a system.
Our vision is to demystify utility functions by providing human stakeholders with lightweight and understandable methods for the definition and refinement of utility functions.

\vspace{0.2em}
\noindent\textbf{Acknowledgments:}
This work is supported in part by the Wallenberg AI, Autonomous Systems and Software Program (WASP) funded by the Knut and Alice Wallenberg Foundation, by award N00014172899 from the Office of Naval Research and by the NSA under Award No.~H9823018D000.
Any views, opinions, findings and conclusions or recommendations expressed in this material are those of the authors and do not necessarily reflect the views of the Office of Naval Research or the NSA.
\vspace{-1.2em}

\end{document}